\documentclass[preprint,12pt,authoryear]{elsarticle}

\usepackage{amssymb}
\usepackage{amsmath}
\usepackage{array}

\journal{Cognitive Systems Research}

\begin{document}

\begin{frontmatter}



\title{Biologically Inspired Deep Learning Approaches for Fetal Ultrasound Image Classification} 


\author[sk,uae]{Rinat Prochii}
\author[sk,uae]{Elizaveta Dakhova}
\author[sk,uae]{Pavel Birulin}
\author[sk,uae]{Maxim Sharaev}

\affiliation[sk]{organization={Skolkovo Institute of Science and Technology},
            city={Moscow},
            postcode={121205},
            country={Russia}}

\affiliation[uae]{organization={BIMAI-Lab, Biomedically Informed Artificial Intelligence Laboratory, University of Sharjah},
            city={Sharjah},
            country={United Arab Emirates}}

\begin{abstract}
Accurate classification of second-trimester fetal ultrasound images remains challenging due to low image quality, high intra-class variability, and significant class imbalance. In this work, we introduce a simple yet powerful, biologically inspired deep learning ensemble framework that—unlike prior studies focused on only a handful of anatomical targets—simultaneously distinguishes 16 fetal structures. Drawing on the hierarchical, modular organization of biological vision systems, our model stacks two complementary branches (a “shallow” path for coarse, low-resolution cues and a “detailed” path for fine, high-resolution features), concatenating their outputs for final prediction.
To our knowledge, no existing method has addressed such a large number of classes with a comparably lightweight architecture. We trained and evaluated on 5,298 routinely acquired clinical images (annotated by three experts and reconciled via Dawid–Skene), reflecting real-world noise and variability rather than a “cleaned” dataset. Despite this complexity, our ensemble (EfficientNet-B0 + EfficientNet-B6 with LDAM-Focal loss) identifies 90\% of organs with accuracy > 0.75 and 75\% of organs with accuracy > 0.85—performance competitive with more elaborate models applied to far fewer categories. These results demonstrate that biologically inspired modular stacking can yield robust, scalable fetal anatomy recognition in challenging clinical settings.
\end{abstract}

\begin{highlights}
\item Best ensemble model achieved 85\% overall accuracy and 0.86 F1‐score, with 55\% accuracy on the most challenging “Umbilical Cord (Anterior Abdominal Wall)” class, matching human expert agreement (54\%).
\item Curated and annotated dataset of 5 298 second‐trimester fetal ultrasound images spanning 16 anatomical classes, with consensus labels obtained from three clinicians via the Dawid–Skene algorithm.
\item On‐the‐fly augmentation pipeline (gamma correction, random crop/resize, flips, color jitter, grayscale, Gaussian blur, translation) to simulate real‐world imaging variability and address severe class imbalance.
\item Novel ensemble architecture combining a “shallow” branch (EfficientNet-B0) for coarse features and a “detailed” branch (EfficientNet-B6) for high‐resolution features, yielding robust feature fusion.
\item Evaluation of imbalance‐aware loss functions (cross‐entropy with label smoothing, focal loss, LDAM, LDAM-Focal), identifying LDAM-Focal as most effective for boosting performance on underrepresented classes.
\end{highlights}


\end{frontmatter}



\section{Introduction}

Ultrasound imaging has become an essential diagnostic technique in prenatal care due to its non‐invasive nature, real‐time capabilities, and absence of ionizing radiation, making it safe for both the mother and fetus. It provides clinicians invaluable insights into fetal anatomy and physiology, facilitating crucial evaluations of fetal health and guiding clinical interventions.

Despite its numerous advantages, fetal ultrasound interpretation — particularly during the second trimester—is subject to considerable challenges. Unlike medical imaging methods that typically focus on a single isolated anatomical structure, fetal ultrasound involves a complex, layered scenario, where ultrasound waves must penetrate maternal tissues to visualize fetal structures. This layering results in image degradation—characterized by signal scattering, attenuation, and acoustic shadowing—which can markedly reduce diagnostic accuracy \citep{dashe2009effect}. Moreover, incomplete surveys are not uncommon when fetal position or maternal habitus impede visualization, with one large study reporting that approximately 4\% of second‐trimester scans remain incomplete due to unfavorable fetal orientation \citep{padula2015rate}.

Ultrasound images typically present two‐dimensional (2D) slices, even though fetal structures are inherently three‐dimensional (3D). Consequently, clinicians must mentally reconstruct complex 3D anatomy from 2D sections, increasing the likelihood of interpretation errors \citep{gonccalves2005three}. Furthermore, ultrasound images are inherently noisy and possess comparatively low resolution \citep{patil2022comparative} which further complicates precise identification of subtle anatomical landmarks critical for accurate fetal assessment.

\begin{figure}[h!]
    \centering
    \includegraphics[width=1\textwidth]{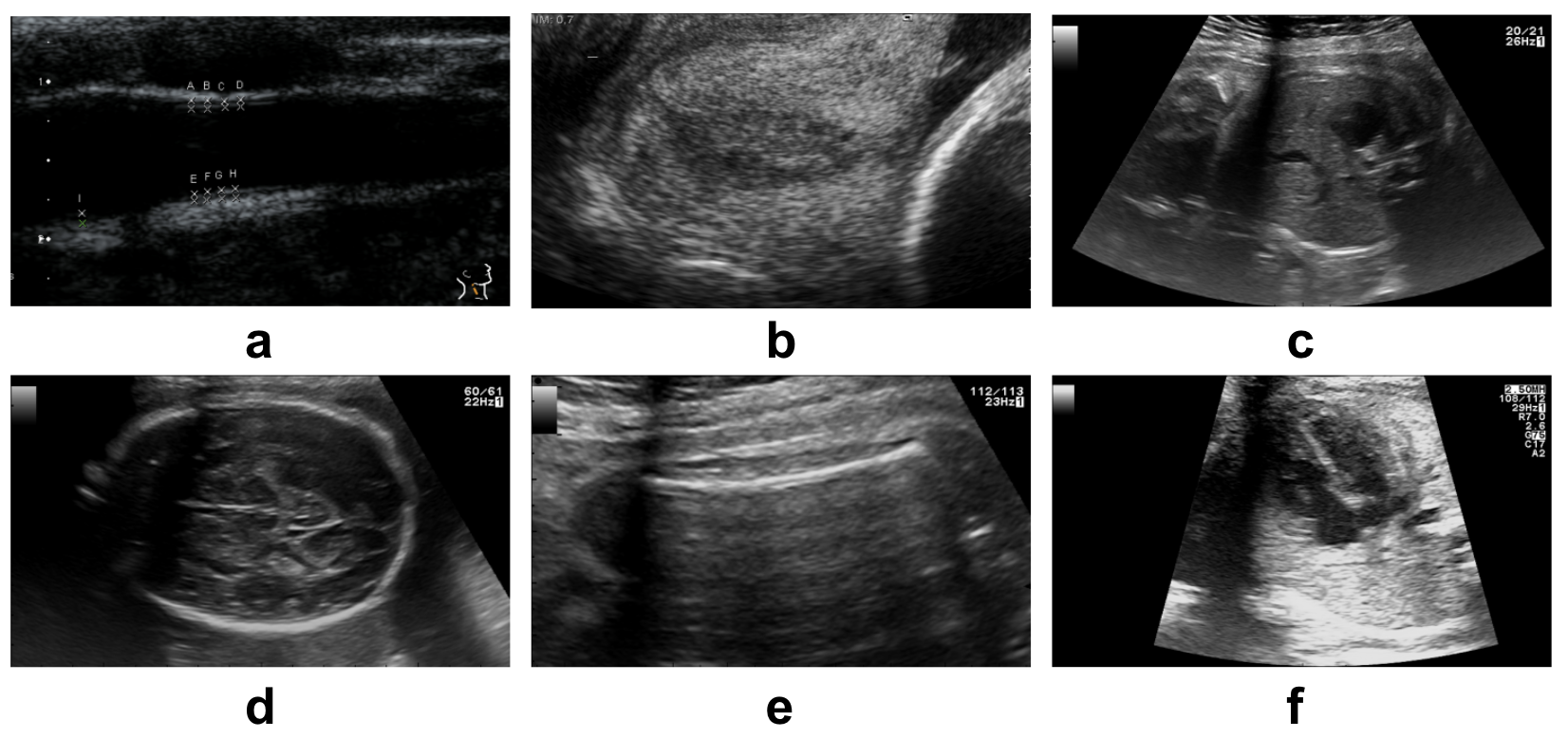}
    \caption{Representative ultrasound images from the Zenodo US Dataset \citep{burgos2020evaluation}. Panels show: (a) Other; (b) Maternal cervix; (c) Fetal abdomen; (d) Fetal brain; (e) Fetal femur; (f) Fetal thorax.}
    \label{fig:zenodo_planes}
\end{figure}

To illustrate the diversity of imaging planes and real‐world variability encountered in prenatal sonography, we reference the Zenodo US Dataset \citep{burgos2020evaluation}. This publicly available repository comprises over 12400 two‐dimensional maternal–fetal ultrasound images collected prospectively from 1792 patients across multiple clinical sites and ultrasound systems. It encompasses six widely used anatomical views. Figure \ref{fig:zenodo_planes} presents representative examples from each of these categories.

Moreover, frequent fetal movements and shifting positions complicate image acquisition, often leading to variable orientations or partial obscuring of critical anatomical structures \citep{baumgartner2017sononet}. Such dynamics introduce subjectivity and inconsistencies into manual image interpretation, as sonographers may need multiple attempts and varying imaging angles to fully evaluate fetal anatomy.

Interpretation accuracy heavily depends on clinician expertise, experience, and training. Even skilled practitioners can face difficulties consistently identifying subtle anatomical details, leading to diagnostic variability \citep{sarris2012intra}. A recent scoping review highlights that current ultrasound training curricula vary widely across institutions, emphasizing the need for standardized programs to reduce operator‐dependent discrepancies \citep{matschl2024current}. Limited image quality—along with class imbalance and underrepresentation of certain fetal structures in clinical datasets—further exacerbates these challenges. A promising approach to robust segmentation and classification of ultrasound images could incorporate concepts from topological data analysis, such as persistent homology for shape characterization \citep{bernstein_topological_2020}.

Recent advances in artificial intelligence (AI) clinical applications, particularly in medical computer vision and deep learning, offer promising solutions to these clinical challenges. Methods for detecting even subtle tissue abnormalities—in both 2D \citep{velichkovsky_convolutional_2021} and 3D \citep{alsahanova_knowledge-informed_2025} are now being extensively developed.  Moreover, deep neural networks have been shown to automatically recognize standard diagnostic planes in real time, guiding operators to acquire correct views even in suboptimal imaging conditions \citep{chen2015standard}. AI‐based super‐resolution techniques can enhance image clarity in low‐resource settings, mitigating noise and resolution issues that often hinder interpretation \citep{boumeridja2025enhancing}. An AI‐driven decision‐support system can significantly reduce variability in manual assessments by providing automated, objective, and reproducible evaluations of fetal structures \citep{xiao2023application}. The challenge of limited labelled data is now being addressed through weakly supervised learning methods \citep{pavlov_weakly_2019}. Nevertheless, achieving robust model performance in real clinical settings remains challenging due to dataset heterogeneity, class imbalance, image noise, and varying clinical conditions \citep{fiorentino2023review}.

\section{Methodology}

Prior to detailing each component, we outline here the overall methodology employed to develop a robust, clinically applicable ultrasound image classification system. First, we describe the dataset assembly and annotation process, including image acquisition, conversion, and multi‐expert labeling. Next, we explain how the raw images were formatted and augmented to mitigate class imbalance and simulate real‐world variability. We then introduce the ensemble deep‐learning framework, which combines a lightweight “shallow” branch (for coarse, low‐resolution feature extraction) and a “detailed” branch (for fine‐grained, high‐resolution feature extraction). Following this, we discuss the selection and implementation of specialized loss functions designed to counteract the pronounced class imbalance. Finally, we present the evaluation strategy, emphasizing the importance of both overall metrics (accuracy, F1‐score) and class‐specific performance as captured by the confusion matrix, with particular attention to the most challenging “Umbilical Cord (Anterior Abdominal Wall)” category. This structured approach ensures that each stage—from data preparation to model assessment—is aligned with the goal of achieving balanced, reliable classification in a clinical context.

\subsection{Dataset Description}

The original ultrasound images were obtained in DICOM format (YBR color space) using advanced ultrasound imaging equipment from General Electric (GE), including GE Voluson E10, GE Voluson E8, and GE Voluson E6 and converted to JPEG (RGB) to ensure compatibility with the annotation platform. Three independent clinicians performed annotations. To reconcile differences among annotators, the Dawid–Skene \citep{dawid1979maximum} method was applied, yielding a single consensus label per image.

A total of 5 298 images spanning 16 classes were initially collected. The full list of classes is presented in table \ref{tab:class_distr}.

\begin{table}[ht!]
    \centering
    \renewcommand{\arraystretch}{1.3} 
    \setlength{\tabcolsep}{3pt} 
    \begin{tabular}{c c}
        \textbf{Class Label} & \textbf{Count} \\ \hline
        Other & 1798 \\ \hline
        Head (PPP, quadrigeminal plate) & 534 \\ \hline
        Four-chamber heart section & 181 \\ \hline
        Section through three vessels & 272 \\ \hline
        Kidneys & 294 \\ \hline
        Stomach & 272 \\ \hline
        Head (sagittal) & 276 \\ \hline
        Head (cerebellum) & 260 \\ \hline
        Umbilical cord (placenta) & 249 \\ \hline
        Spine & 192 \\ \hline
        Femur & 208 \\ \hline
        Nasolabial triangle & 170 \\ \hline
        Humerus & 173 \\ \hline
        Umbilical cord (anterior abdominal wall) & 149 \\ \hline
        Cervix & 143 \\ \hline
        Bladder (CDC) & 138 \\ \hline
        Overall & 5298 
    \end{tabular}
    \caption{Classes Distribution}
    \label{tab:class_distr}
\end{table}

The class distribution is highly imbalanced, which presents challenges for model training and necessitates careful handling during preprocessing and augmentation.

\subsubsection{Data Augmentation}

To mitigate class imbalance and improve model robustness, the following augmentations were applied on the fly during training:
\begin{itemize}
    \item \textbf{Gamma Correction}: Adjusts image intensity via $I_{\text{out}}(x,y)=I_{\text{in}}(x,y)^\gamma$, where $\gamma$ is sampled from a predefined range.
    \item \textbf{Random Crop and Resize}: A random subregion is cropped, then resized to the target dimensions using bilinear interpolation.
    \item \textbf{Horizontal/Vertical Flip}: Random flips to introduce symmetry invariance.
    \item \textbf{Color Jitter}: Randomly perturbs brightness, contrast, saturation, and hue within specified bounds.
    \item \textbf{Grayscale Conversion}: Converts RGB images to a single‐channel intensity map.
    \item \textbf{Gaussian Blur}: Applies a Gaussian kernel $\displaystyle G(u,v)=\frac{1}{2\pi\sigma^2}\exp\bigl(-\tfrac{u^2+v^2}{2\sigma^2}\bigr)$ to smooth image details.
    \item \textbf{Translation}: Randomly shifts the image by $(\Delta x,\Delta y)$ pixels, with out‐of‐bounds regions zero‐padded.
\end{itemize}

These augmentations introduce realistic variability (e.g., lighting changes, cropping variations), helping to prevent overfitting and improving generalization under different acquisition conditions.

\subsection{Deep Learning Models}

Given the inherent challenges posed by the low quality and variability of ultrasound images, an ensemble approach of deep learning models, inspired by brain hierarchical architecture, was adopted as the primary classification strategy. This ensemble framework enhances robustness and overall performance by integrating features extracted at different levels of detail, thereby improving the model's ability to capture both global and fine-grained information in the images.

The ensemble is comprised of two primary branches:
\begin{itemize}
    \item \textbf{Shallow Model:} A lightweight network designed to capture coarse, lower-resolution features that are critical for establishing a general understanding of the image structure.
    \item \textbf{Detailed Model:} A deeper and more complex network tailored to extract high-resolution, detailed features essential for discriminating subtle image characteristics.
\end{itemize}

The overall architecture processes an input image by considering it at two distinct scales and subsequently fusing the features obtained from each branch for final classification. The forward pass through the ensemble network can be summarized as follows (see Figure \ref{fig:ensemble}):
 
\begin{figure}[h!]
    \centering
    \includegraphics[width=1\textwidth]{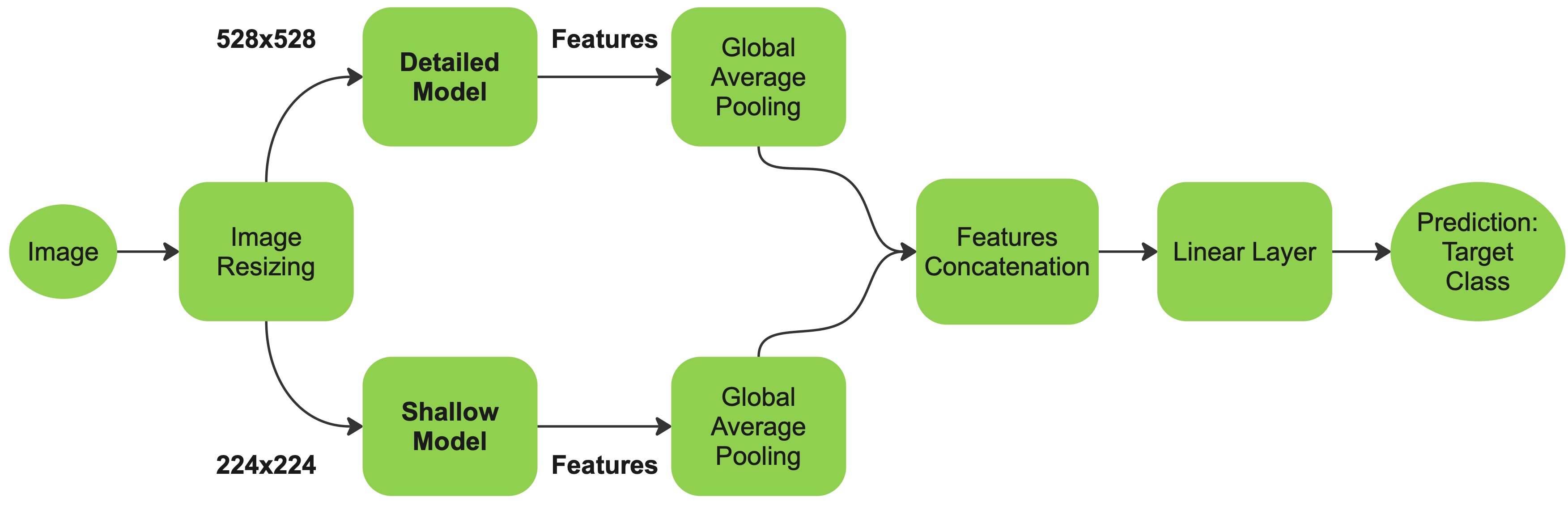}
    \caption{Biologically Inspired Ensemble Architecture Overview. The input image is resized to two scales—a low‐resolution version ($x_s$) for the shallow backbone and a high‐resolution version ($x_d$) for the detailed backbone. Each branch extracts feature maps that are then reduced via global average pooling and concatenated into a single feature vector for final classification.}
    \label{fig:ensemble}
\end{figure}

\begin{enumerate}
    \item \textbf{Resizing the Image:}  
    The input image $x$ is first rescaled to two different sizes. In some cases, scaling factors are applied rather than fixed dimensional rescaling.:
    \begin{itemize}
        \item $x_s$: A smaller-scale version (e.g., approximately $224\times224$ pixels), optimized for the shallow model. 
        \item $x_d$: A larger-scale version (e.g., approximately $528\times528$ pixels), which preserves more detail for the detailed model.
    \end{itemize}

    \item \textbf{Feature Extraction:}  
    The shallow backbone processes $x_s$ to extract broader, low-resolution features, while the detailed backbone processes $x_d$ to extract rich, high-resolution feature maps.

    \item \textbf{Global Average Pooling (GAP):}  
    The feature maps generated by each network are subjected to adaptive average pooling to reduce the spatial dimensions to a fixed size (typically $1 \times 1$). This is mathematically represented as:
    \begin{equation}
        GAP(f) = \frac{1}{H\times W}\sum_{i=1}^H \sum_{j=1}^W f_{i,j},
    \end{equation}
    where $H \times W$ corresponds to the height and width of the feature map.

    \item \textbf{Feature Fusion:}  
    The pooled feature vectors from the shallow and detailed models are concatenated along the channel dimension to form a comprehensive feature representation that is then fed into subsequent classification layers.
\end{enumerate}

In practice, different ResNet \citep{he2016deep} and EfficientNet \citep{tan2019efficientnet} architectures have been experimented with for each branch to identify the optimal combination, including  ResNet-34 and EfficientNet-B6 for the detailed model, as well as EfficientNet-B0, ResNet-18 for the shallow model. This ensemble design allows the model to effectively adapt to real-world clinical ultrasound images by maintaining robustness even in scenarios with significant variation in image quality.

\subsection{Loss Functions}

To address the challenges associated with a highly imbalanced dataset, multiple loss functions were investigated. These loss functions dynamically adjust the computation of the loss to improve the classification performance for underrepresented classes. The following loss functions were evaluated:

\begin{itemize}
    \item Cross-Entropy Loss
    \item Cross-Entropy Loss with Label Smoothing
    \item Focal Loss \citep{ross2017focal}
    \item Label-Distribution-Aware Margin (LDAM) Loss \citep{cao2019learning}
    \item LDAM-Focal Loss \citep{sadi2022lmfloss}
\end{itemize}

\subsection{Metrics}

Two primary metrics were employed for evaluation: overall accuracy and F1‐score. However, the most informative assessment was derived from the confusion matrix, which provided class‐specific accuracy values. Because the model was being developed for clinical application, the primary objective was to ensure balanced performance across all classes. Accordingly, we visually examined the per‐class accuracies in the confusion matrix to assess and verify that the model achieved equitable accuracy for each category.

\begin{equation}
    \mathrm{Accuracy} = \frac{TP + TN}{TP + TN + FP + FN}
\end{equation}

\begin{equation}
    \mathrm{Precision} = \frac{TP}{TP + FP}
\end{equation}

\begin{equation}
    \mathrm{Recall} = \frac{TP}{TP + FN}
\end{equation}

\begin{equation}
    \text{F1-Score} = 2 \cdot \frac{\mathrm{Precision} \times \mathrm{Recall}}{\mathrm{Precision} + \mathrm{Recall}}
\end{equation}

where $TP,\ TN,\ FP,\ FN$ denote true positives, true negatives, false positives, and false negatives, respectively.

Among all categories, the “Umbilical Cord (Anterior Abdominal Wall (AAW))” class—which denotes the point where the umbilical cord attaches to the fetal anterior abdominal wall—was the most difficult to classify. Although the model exhibited high overall accuracy and F1-score, we specifically aimed to maximize accuracy for this particular class. As a result, optimizing for the “Umbilical Cord (AAW)” label required a deliberate trade-off during model tuning.

\section{Results}

We performed approximately sixty experiments, varying hyperparameters and loss functions, to identify the most effective ensemble configurations. Table \ref{tab:metrics} summarizes the top ten results, all trained with the Adam optimizer \citep{kingma2014adam}. Because our primary objective was to maximize both overall performance and accuracy on the “Umbilical Cord (Anterior Abdominal Wall)” class, the ensemble combining EfficientNet-B0 and EfficientNet-B6 with LDAM-Focal Loss emerged as the best solution.

\begin{table}[ht!]
    \centering
    \renewcommand{\arraystretch}{1.3} 
    \setlength{\tabcolsep}{3pt} 
    \begin{tabular}{
        p{3.5cm}
        p{4cm}
        >{\centering\arraybackslash}p{2cm}
        c
        >{\centering\arraybackslash}p{2cm}
    }
        \textbf{Ensemble Architecture} & \textbf{Loss Function} & \textbf{Accuracy Score} & \textbf{F1-Score} & \textbf{Umbilical Cord (AAW) Accuracy} \\ \hline
        ResNet18+ ResNet34 & CrossEntropy Loss & 85\% & 0.84 & 21\% \\ \hline
        ResNet18+ ResNet34 & CrossEntropy Loss with Label Smoothing & 86\% & 0.85 & 21\% \\ \hline
        ResNet18+ ResNet34 & Focal Loss & 83\% & 0.83 & 18\% \\ \hline
        ResNet18+ ResNet34 & LDAM Loss & 82\% & 0.82 & 24\% \\ \hline
        ResNet18+ ResNet34 & LDAM-Focal Loss & 81\% & 0.81 & 33\% \\ \hline
        EfficientNet-B0+ EfficientNet-B6 & CrossEntropy Loss & 86\% & 0.86 & 42\% \\ \hline
        EfficientNet-B0+ EfficientNet-B6 & CrossEntropy Loss with Label Smoothing & \textbf{87\%} & \textbf{0.87} & 39\% \\ \hline
        EfficientNet-B0+ EfficientNet-B6 & Focal Loss & 86\% & 0.86 & 33\% \\ \hline
        EfficientNet-B0+ EfficientNet-B6 & LDAM Loss & 84\% & 0.84 & 45\% \\ \hline
        EfficientNet-B0+ EfficientNet-B6 & LDAM-Focal Loss & 85\% & 0.86 & \textbf{55\%} \\ \hline 
    \end{tabular}
    \caption{Ensemble Metrics}
    \label{tab:metrics}
\end{table}

Table \ref{tab:per_class_accuracy} displays per‐class accuracies. Most anatomical categories achieved classification accuracies exceeding 80\%. However, the exceptions include the \textit{Kidneys} (77\% accuracy), \textit{Placenta (Umbilical Cord)} (70\% accuracy), and \textit{Four-chamber heart section} (78\% accuracy), each remaining marginally below the 80\% accuracy threshold.

\begin{table}[ht!]
    \centering
    \renewcommand{\arraystretch}{1.3} 
    \setlength{\tabcolsep}{3pt} 
    \begin{tabular}{p{8cm} >{\centering\arraybackslash}p{3cm}}
        \textbf{Class Label} & \textbf{Accuracy (\%)} \\ \hline
        Femur & 94 \\ \hline
        Head (PPP, Tectum) & 91 \\ \hline
        Head (Cerebellum) & 90 \\ \hline
        Head (Sagittal) & 95 \\ \hline
        Other & 84 \\ \hline
        Stomach & 88 \\ \hline
        Bladder (CDC) & 85 \\ \hline
        Nasal triangle & 84 \\ \hline
        Shoulder bone & 96 \\ \hline
        Spine & 85 \\ \hline
        Kidneys & 77 \\ \hline
        Umbilical cord (Anterior abdominal wall) & 55 \\ \hline
        Placenta (Umbilical cord) & 70 \\ \hline
        Slice through three vessels & 86 \\ \hline
        Four-chamber heart section & 78 \\ \hline
        Cervix & 93 \\ \hline
    \end{tabular}
    \caption{Per‐class accuracies extracted from the main diagonal of the confusion matrix.}
    \label{tab:per_class_accuracy}
\end{table}

\section{Discussion}

In this work, we set out to overcome the key obstacles in second‐trimester fetal ultrasound classification—namely, low image quality, high intra‐class variability, severe class imbalance, and the unprecedented scale of 16 anatomical categories—by adopting a biologically inspired, hierarchical ensemble design. Leveraging a dual‐branch architecture that mirrors the coarse‐to‐fine processing observed in biological vision systems, our model achieved an overall accuracy of 85\% and an F1‐score of 0.86, while identifying 90\% of organs with accuracy > 0.75 and 75\% with accuracy > 0.85. Notably, per‐class accuracies exceeded 80\% for 12 out of 16 structures—levels of performance that, to our knowledge, have only been attained previously in studies addressing far fewer classes.

Despite this strong overall performance, several anatomies remained more challenging. The \textit{Kidneys} (77\% accuracy), \textit{Four-chamber heart section} (78\%), and \textit{Placenta (Umbilical Cord)} (70\%) illustrate how overlapping echogenic signatures and variable fetal orientations can still impede consistent detection—issues first highlighted in our Introduction when discussing signal scattering, acoustic shadowing, and incomplete surveys in up to 4\% of clinical scans. Yet even here, our simple stacking of two complementary backbones delivered meaningful improvements over both single‐model baselines and prior works limited to small class sets.

One particular anatomical class, the \textit{Umbilical Cord (Anterior Abdominal Wall)}, merits additional discussion. Its overall accuracy (55\%) remained comparatively low. The challenge with recognizing the \textit{Umbilical Cord (Anterior Abdominal Wall)} arises primarily due to inherent anatomical ambiguity; this structure often co-occurs as a secondary anatomical element in ultrasound images labeled predominantly with other primary classes. Such anatomical overlap complicates accurate identification and labeling.

To further investigate this ambiguity, we conducted a supplementary labeling experiment involving two clinical experts. Both experts were independently tasked with annotating a consistent set of images in which the \textit{Umbilical Cord (Anterior Abdominal Wall)} could potentially be present. Notably, Expert A identified 880 images as positively containing this anatomical feature, whereas Expert B marked only 455 images. Crucially, only 492 annotations overlapped between the experts, representing approximately 54\% agreement. Remarkably, this inter-rater agreement closely mirrors the accuracy (55\%) achieved by developed model, effectively matching human-level consistency. This finding underscores the inherent difficulty and subjectivity in recognizing this particular anatomical feature and indicates that the developed model has effectively reached the practical accuracy limit determined by human expert consensus.

Taken together, these findings validate the value of biologically inspired modular stacking: by first capturing coarse, low‐resolution context and then refining with high‐resolution detail, the ensemble robustly navigates real‐world noise and class imbalance across an unprecedented number of targets. This approach not only addresses the specific challenges outlined in our Introduction but also offers a scalable paradigm for future expansions—be it additional classes, imaging devices, or incorporation of anatomical priors.

\section{Conclusion}

This work presents a clinically oriented deep learning–based classification system for second‐trimester fetal ultrasound images that addresses low image quality, class imbalance, and inter‐class ambiguity. By leveraging a dual‐branch ensemble architecture (EfficientNet-B0 + EfficientNet-B6) trained with LDAM-Focal Loss, we achieved high overall F1-Score (86\%) and balanced per‐class performance. Crucially, our model’s 55\% accuracy on the “Umbilical Cord (Anterior Abdominal Wall)” class aligns closely with human expert consensus (54\%), indicating that further gains in this specific category may be constrained by annotation ambiguity rather than model capacity.

These findings underscore the effectiveness of combining coarse and fine feature extraction for complex ultrasound classification tasks and highlight the importance of using imbalance‐aware loss formulations. By demonstrating human‐level consistency on the most difficult class and achieving > 80\% accuracy on most categories, this framework shows promise for real‐time decision‐support in prenatal care. Future work will explore (1) expanding the dataset to include additional ultrasound devices and clinical sites to improve generalizability, (2) integrating domain‐specific anatomical priors (e.g., spatial relationships between fetal structures) to further reduce ambiguity, and (3) conducting prospective clinical trials to assess the impact of this AI‐driven system on diagnostic workflow efficiency and inter‐operator variability.



\bibliographystyle{elsarticle-harv}
\bibliography{main.bib}






\end{document}